\documentclass[aps,twocolumn,twoside,bibnotes,preprintnumbers]{revtex4}
\usepackage{graphicx,amsfonts}

\newcommand{\<}{\langle}
\renewcommand{\>}{\rangle}

\newcommand{\be}{\begin{equation}}
\newcommand{\ee}{\end{equation}}
\newcommand{\bea}{\begin{eqnarray}}
\newcommand{\eea}{\end{eqnarray}}

\newcommand{\tr}{\mathop{\mathrm{tr}}}
\newcommand{\poly}{\mathop{\mathrm{poly}}}

\newcommand{\norm}[1]{\left\|#1\right\|}

\begin{document}

\title{Quantum search by measurement}

\author{Andrew M. Childs}
\email[]{amchilds@mit.edu}
\affiliation{Center for Theoretical Physics,
             Massachusetts Institute of Technology,
	     Cambridge, MA 02139, USA} 

\author{Enrico Deotto}
\email[]{deotto@mitlns.mit.edu}
\affiliation{Center for Theoretical Physics,
             Massachusetts Institute of Technology,
	     Cambridge, MA 02139, USA} 

\author{Edward Farhi}
\email[]{farhi@mit.edu}
\affiliation{Center for Theoretical Physics,
             Massachusetts Institute of Technology,
	     Cambridge, MA 02139, USA} 

\author{Jeffrey Goldstone}
\email[]{goldston@mit.edu}
\affiliation{Center for Theoretical Physics,
             Massachusetts Institute of Technology,
	     Cambridge, MA 02139, USA} 

\author{Sam Gutmann}
\email[]{sgutm@neu.edu}
\affiliation{Department of Mathematics,
             Northeastern University,
	     Boston, MA 02115, USA}

\author{Andrew J. Landahl}
\email[]{alandahl@caltech.edu}
\affiliation{Institute for Quantum Information,
             California Institute of Technology,
	     Pasadena, CA 91125, USA}

\date[]{3 April 2002}

\preprint{MIT-CTP \#3260}


\begin{abstract}
We propose a quantum algorithm for solving combinatorial search problems
that uses only a sequence of measurements.  The algorithm is similar in
spirit to quantum computation by adiabatic evolution, in that the goal is
to remain in the ground state of a time-varying Hamiltonian.  Indeed, we
show that the running times of the two algorithms are closely related.  We
also show how to achieve the quadratic speedup for Grover's unstructured
search problem with only two measurements.  Finally, we discuss some
similarities and differences between the adiabatic and measurement
algorithms.
\end{abstract}

\maketitle

\section{Introduction}

In the conventional circuit model of quantum computation, a program for a
quantum computer consists of a discrete sequence of unitary gates chosen
from a fixed set.  The memory of the quantum computer is a collection of
qubits initially prepared in some definite state.  After a sequence of
unitary gates is applied, the qubits are measured in the computational
basis to give the result of the computation, a classical bit string.

This description of a quantum computer has been used to formulate quantum
algorithms that outperform classical methods, notably Shor's factoring
algorithm \cite{Sho94} and Grover's algorithm for unstructured search
\cite{Gro97}.  Subsequent development of quantum algorithms has focused
primarily on variations of the techniques introduced by Shor and Grover.
One way to motivate new algorithmic ideas is to consider alternative (but
in general, equivalent) descriptions of the way a quantum computer
operates.  For example, the technique of quantum computation by adiabatic
evolution \cite{FGGS00} is most easily described by a quantum computer
that evolves continuously according to a time-varying Hamiltonian.

Another model of quantum computation allows measurement at intermediate
stages.  Indeed, recent work has shown that {\em measurement alone} is
universal for quantum computation: one can efficiently implement a
universal set of quantum gates using only measurements (and classical
processing) \cite{Nie01, FZ01, Leu01, RB00}.  In this paper, we describe
an algorithm for solving combinatorial search problems that consists only
of a sequence of measurements.  Using a straightforward variant of the
quantum Zeno effect (see for example \cite{Neu32, AV80, SRM93}), we show
how to keep the quantum computer in the ground state of a smoothly varying
Hamiltonian $H(s)$.  This process can be used to solve a computational
problem by encoding the solution to the problem in the ground state of the
final Hamiltonian.

The organization of the paper is as follows.  In Section
\ref{sec:algorithm}, we present the algorithm in detail and describe how
measurement of $H(s)$ can be performed on a digital quantum computer.  In
Section \ref{sec:runtime}, we estimate the running time of the algorithm
in terms of spectral properties of $H(s)$.  Then, in Section
\ref{sec:grover}, we discuss how the algorithm performs on Grover's
unstructured search problem and show that by a suitable modification,
Grover's quadratic speedup can be achieved by the measurement algorithm.
Finally, in Section \ref{sec:discussion}, we discuss the relationship
between the measurement algorithm and quantum computation by adiabatic
evolution.

\section{The measurement algorithm}
\label{sec:algorithm}

Our algorithm is conceptually similar to quantum computation by adiabatic
evolution \cite{FGGS00}, a general method for solving combinatorial
search problems using a quantum computer.  Both algorithms operate by
remaining in the ground state of a smoothly varying Hamiltonian $H(s)$
whose initial ground state is easy to construct and whose final ground
state encodes the solution to the problem.  However, whereas adiabatic
quantum computation uses Schr\"odinger evolution under $H(s)$ to remain in
the ground state, the present algorithm uses {\em only} measurement of
$H(s)$.

In general, we are interested in searching for the minimum of a function
$h(z)$ that maps $n$-bit strings to positive real numbers.  Many
computational problems can be cast as minimization of such a function; for
specific examples and their relationship to adiabatic quantum computation,
see \cite{FGGS00, CFGG01}.  Typically, we can restrict our attention to
the case where the global minimum of $h(z)$ is unique.  Associated with
this function, we can define a {\em problem Hamiltonian} $H_P$ through its
action on computational basis states:
\be
  H_P |z\> = h(z) |z\>
\,.
\ee
Finding the global minimum of $h(z)$ is equivalent to finding the ground
state of $H_P$.  If the global minimum is unique, then this ground state
is nondegenerate.

To reach the ground state of $H_P$, we begin with the quantum computer
prepared in the ground state of some other Hamiltonian $H_B$, the {\em
beginning Hamiltonian}.  Then we consider a one-parameter family of
Hamiltonians $H(s)$ that interpolates smoothly from $H_B$ to $H_P$ for $s
\in [0,1]$.  In other words, $H(0)=H_B$ and $H(1)=H_P$, and the
intermediate $H(s)$ is a smooth function of $s$.  One possible choice is
linear interpolation,
\be
  H(s) = (1-s) H_B + s H_P
\,.
\label{eq:lininterp}
\ee

Now we divide the interval $[0,1]$ into $M$ subintervals of width $\delta
= 1/M$.  So long as the interpolating Hamiltonian $H(s)$ is smoothly
varying and $\delta$ is small, the ground state of $H(s)$ will be close to
the ground state of $H(s+\delta)$.  Thus, if the system is in the ground
state of $H(s)$ and we measure $H(s+\delta)$, the post-measurement state
is very likely to be the ground state of $H(s+\delta)$.  If we begin in
the ground state of $H(0)$ and successively measure $H(\delta),
H(2\delta), \ldots, H((M-1)\delta), H(1)$, then the final state will be
the ground state of $H(1)$ with high probability, assuming $\delta$ is
sufficiently small.

To complete our description of the quantum algorithm, we must explain how
to measure the operator $H(s)$.  The technique we use is motivated by von
Neumann's description of the measurement process \cite{Neu32}.  In this
description, measurement is performed by coupling the system of interest
to an ancillary system, which we call the {\em pointer}.  Suppose that the
pointer is a one-dimensional free particle and that the system-pointer
interaction Hamiltonian is $H(s) \otimes p$, where $p$ is the momentum of
the particle.  Furthermore, suppose that the mass of the particle is
sufficiently large that we can neglect the kinetic term.  Then the
resulting evolution is
\be
  e^{-i t H(s) \otimes p} 
  = \sum_a \left[ |E_a(s)\>\<E_a(s)| \otimes e^{-i t E_a(s) p} \right]
\,,
\label{eq:measureevolve}
\ee
where $|E_a(s)\>$ are the eigenstates of $H(s)$ with eigenvalues $E_a(s)$,
and we have set $\hbar = 1$.  Suppose we prepare the pointer in the state
$|x=0\>$, a narrow wave packet centered at $x=0$.  Since the momentum
operator generates translations in position, the above evolution performs
the transformation
\be
  |E_a(s)\> \otimes |x=0\> \to |E_a(s)\> \otimes |x= t E_a(s)\>
\,.
\label{eq:translate}
\ee
If we can measure the position of the pointer with sufficiently high
precision that all relevant spacings $x_{ab} = t|E_a(s)-E_b(s)|$ can be
resolved, then measurement of the position of the pointer --- a fixed,
easy-to-measure observable, independent of $H(s)$ --- effects a
measurement of $H(s)$.

Von Neumann's measurement protocol makes use of a continuous variable, the
position of the pointer.  To turn it into an algorithm that can be
implemented on a fully digital quantum computer, we can approximate the
evolution (\ref{eq:measureevolve}) using $r$ quantum bits to represent the
pointer \cite{Wie96, Zal98}.  The full Hilbert space is thus a tensor
product of a $2^n$-dimensional space for the system and a
$2^r$-dimensional space for the pointer.  We let the computational basis
of the pointer, with basis states $\{|z\>\}$, represent the basis of
momentum eigenstates.  The label $z$ is an integer between $0$ and
$2^r-1$, and the $r$ bits of the binary representation of $z$ specify the
states of the $r$ qubits.  In this basis, the digital representation of
$p$ is
\be
  p = \sum_{j=1}^r 2^{-j} {1-\sigma_z^{(j)} \over 2}
\,,
\label{eq:momentum}
\ee
a sum of diagonal operators, each of which acts on only a single qubit.
Here $\sigma_z^{(j)}$ is the Pauli $z$ operator on the $j$th qubit.  As we
will discuss in the next section, we have chosen to normalize $p$ so that
\be
  p|z\>={z \over 2^r}|z\>
\,,
\label{eq:momentumnorm}
\ee
which gives $\norm p \sim 1$.  If $H(s)$ is a sum of terms, each of which
acts on at most $k$ qubits, then $H(s) \otimes p$ is a sum of terms, each
of which acts on at most $k+1$ qubits.  As long as $k$ is a fixed constant
independent of the problem size $n$, such a Hamiltonian can be simulated
efficiently on a quantum computer \cite{Llo96}.  Expanded in the momentum
eigenbasis, the initial state of the pointer is
\be
  |x=0\> = {1 \over 2^{r/2}} \sum_{z=0}^{2^r-1} |z\>
\,.
\label{eq:pointerstate}
\ee
The measurement is performed by evolving under $H(s) \otimes p$ for a
total time $\tau$.  We discuss how to choose $\tau$ in the next section.
After this evolution, the position of the simulated pointer could be
measured by measuring the qubits that represent it in the $x$ basis, i.e.,
the Fourier transform of the computational basis.  However, note that our
algorithm only makes use of the post-measurement state of the system, not
of the measured value of $H(s)$.  In other words, only the reduced density
matrix of the system is relevant.  Thus it is not actually necessary to
perform a Fourier transform before measuring the pointer, or even to
measure the pointer at all.  When the system-pointer evolution is
finished, one can either re-prepare the pointer in its initial state
$|x=0\>$ or discard it and use a new pointer, and immediately begin the
next measurement.

As an aside, note that the von Neumann measurement procedure described
above is identical to the well-known phase estimation algorithm for
measuring the eigenvalues of a unitary operator \cite{Kit95, CEMM98},
which can also be used to produce eigenvalues and eigenvectors of a
Hamiltonian \cite{AL99}.  This connection has been noted previously in
\cite{Zal98}, and it has been pointed out that the measurement is a
non-demolition measurement in \cite{TMR02}.  In the phase estimation
problem, we are given an eigenvector $|\psi\>$ of a unitary operator $U$
and asked to determine its eigenvalue $e^{-i \phi}$.  The algorithm uses
two registers, one that initially stores $|\psi\>$ and one that will store
an approximation of the phase $\phi$.  The first and last steps of the
algorithm are Fourier transforms on the phase register.  The intervening
step is to perform the transformation
\be
  |\psi\> \otimes |z\> \to U^z |\psi\> \otimes |z\>
\,,
\ee
where $|z\>$ is a computational basis state.  If we take $|z\>$ to be a
momentum eigenstate with eigenvalue $z$ (i.e., if we choose a different
normalization than in (\ref{eq:momentumnorm})) and let $U=e^{-i H t}$,
this is exactly the transformation induced by $e^{-i (H \otimes p) t}$.
Thus we see that the phase estimation algorithm for a unitary operator $U$
is exactly von Neumann's prescription for measuring $i\ln U$.

\section{Running time}
\label{sec:runtime}

The running time of the measurement algorithm is the product of $M$, the
number of measurements, and $\tau$, the time per measurement.  Even if we
assume perfect projective measurements, the algorithm is guaranteed to
keep the computer in the ground state of $H(s)$ only in the limit $M \to
\infty$, so that $\delta = 1/M \to 0$.  Given a finite running time, the
probability of finding the ground state of $H_P$ with the last measurement
will be less than 1.  To understand the efficiency of the algorithm, we
need to determine how long we must run as a function of $n$, the number of
bits on which the function $h$ is defined, so that the probability of
success is not too small.  In general, if the time required to achieve a
success probability greater than some fixed constant (e.g., $1 \over 2$)
is $\poly(n)$, we say the algorithm is efficient, whereas if the running
time grows exponentially, we say it is not.

To determine the running time of the algorithm, we consider the effect of
the measurement process on the reduced density matrix of the system.
Here, we simply motivate the main result; for a detailed analysis, see
Appendix~\ref{app:measurement}.

\begin{figure}
\includegraphics[width=3.3in]{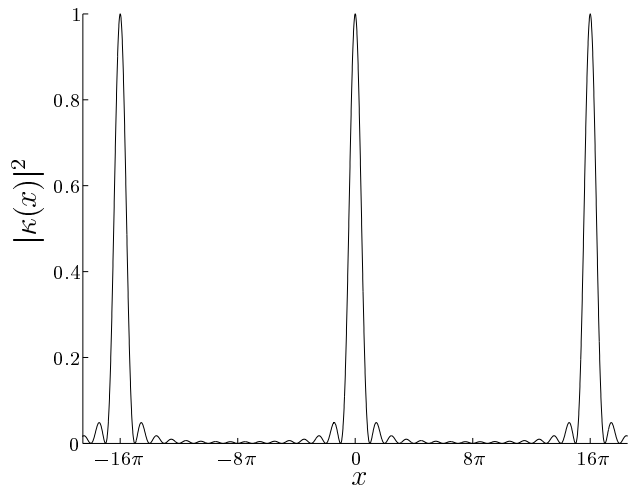}
\caption{The function $|\kappa(x)|^2$ for $r=4$.}
\label{fig:kappa}
\end{figure}

Let $\rho^{(j)}$ denote the reduced density matrix of the system after the
$j$th measurement; its matrix elements are
\be
  \rho^{(j)}_{ab} = \<E_a(j\delta)|\rho^{(j)}|E_b(j\delta)\>
\,.
\ee
The interaction with the digitized pointer effects the transformation
\be
  |E_a(s)\> \otimes |z\> 
    \to e^{-i E_a(s) z t / 2^r} |E_a(s)\> \otimes |z\>
\,.
\label{eq:pointerevolve}
\ee
Starting with the pointer in the state (\ref{eq:pointerstate}), evolving
according to (\ref{eq:pointerevolve}), and tracing over the pointer, the
quantum operation induced on the system is 
\be
  \rho^{(j+1)}_{ab} = \kappa^{(j+1)}_{ab} 
                      \sum_{c,d} U^{(j)}_{ac} \rho^{(j)}_{cd} 
		                 U^{(j)*}_{bd}
\,,
\label{eq:systempointer}
\ee
where the unitary transformation relating the energy eigenbases at
$s=j\delta$ and $s=(j+1)\delta$ is
\be
  U^{(j)}_{ab} = \< E_a((j+1)\delta) | E_b(j\delta) \>
\label{eq:unitary}
\ee
and
\be
  \kappa^{(j)}_{ab} = {1 \over 2^r} \sum_{z=0}^{2^r-1} 
                      e^{i[E_b(j\delta)-E_a(j\delta)]zt/2^r}
\,.
\ee
Summing this geometric series, we find
\be
  \left|\kappa^{(j)}_{ab}\right|^2 
    = \left|\kappa([E_b(j\delta)-E_a(j\delta)]t/2)\right|^2
\,,
\ee
where
\be
  |\kappa(x)|^2 = {\sin^2 x \over 4^r \sin^2(x/2^r)}
\,.
\label{eq:kappa}
\ee
This function is shown in Fig.~\ref{fig:kappa} for the case $r=4$.  It
has a sharp peak of unit height and width of order 1 at the origin, and
identical peaks at integer multiples of $2^r \pi$.

If the above procedure were a perfect projective measurement, then we
would have $\kappa_{ab}=0$ whenever $E_a \ne E_b$.  Assuming (temporarily)
that this is
the case, we find
\be
  \rho^{(j+1)}_{00} \ge \left|U^{(j)}_{00}\right|^2 \rho^{(j)}_{00} 
\label{eq:perfectmeas}
\ee
with the initial condition $\rho^{(0)}_{00}=1$ and $\rho^{(0)}_{ab}=0$
otherwise.  Perturbation theory gives
\bea
  \left|U^{(j)}_{00}\right|^2
  &=&   1 - \delta^2 \sum_{a \ne 0} 
        {|\<E_a(s)|{{\mathrm d}H \over {\mathrm d}s}|E_0(s)\>|^2 \over
        (E_0(s)-E_a(s))^2} \Bigg|_{s=j\delta} \nonumber \\
  &&    + O(\delta^3) \\
  &\ge& 1 - {\Gamma(j\delta)^2 \, \delta^2 \over g(j\delta)^2} + O(\delta^3)
\,,
\label{eq:perturb}
\eea
where
\be
  \Gamma(s)^2
    = \<E_0(s)|({\textstyle{{\mathrm d}H \over {\mathrm d}s}})^2|E_0(s)\>
     -\<E_0(s)| {\textstyle{{\mathrm d}H \over {\mathrm d}s}}|E_0(s)\>^2
\ee
and
\be
  g(s) = E_1(s)-E_0(s)
\ee
is the energy gap between the ground and first excited states.  If we let 
\bea
  \Gamma &=& \max_{s\in[0,1]} \Gamma(s) \\
  g      &=& \min_{s\in[0,1]} g(s)
\,,
\eea
then according to (\ref{eq:perfectmeas}), the probability of being in the
ground state after the last measurement is at least
\bea
  \rho^{(M)}_{00}
    &\ge& \left[1-{\Gamma^2 \over M^2 g^2} 
                 +O(M^{-3}) \right]^M \\
    &=&   \exp\left(-{\Gamma^2 \over M g^2}\right) + O(M^{-2})
\,.
\eea
The probability of success is close to 1 provided
\be
  M \gg {\Gamma^2 \over g^2}
\,.
\label{eq:numbermeas}
\ee
When $H_B$ and $H_P$ are both sums of $\poly(n)$ terms, each of which acts
nontrivially on at most a constant number of qubits, it is easy to choose
an interpolation such as (\ref{eq:lininterp}) so that $\Gamma$ is only
$\poly(n)$.  Thus we are mainly interested in the behavior of $g$, the
{\em minimum gap} between the ground and first excited states.  We see
that for the algorithm to be successful, the total number of measurements
must be much larger than $1/g^2$.

In fact, the simulated von Neumann procedure is not a perfect projective
measurement.  We must determine how long the system and pointer should
interact so that the measurement is sufficiently good.  The analysis in
Appendix~\ref{app:measurement} shows that $|\kappa^{(j)}_{01}|^2$ should
be bounded below 1 by a constant for all $j$.  In other words, to
sufficiently resolve the difference between the ground and first excited
states, we must decrease the coherence between them by a fixed fraction
per measurement.  The width of the central peak in Fig.~\ref{fig:kappa} is
of order 1, so it is straightforward to show that to have $|\kappa(x)|^2$
less than, say, $1/2$, we must have $x \ge O(1)$.  This places a lower
bound on the system-pointer interaction time of
\be
  \tau \ge {O(1) \over g}
\label{eq:meastime}
\ee
independent of $r$, the number of pointer qubits.  (Note that the same
bound also holds in the case of a continuous pointer with a fixed
resolution length.)

Putting these results together, we find that the measurement algorithm is
successful if the total running time, $T = M \tau$, satisfies
\be
  T \gg {\Gamma^2 \over g^3} \quad \textrm{(measurement)}
\,.
\label{eq:measurebound}
\ee
This result can be compared to the corresponding expression for quantum
computation by adiabatic evolution,
\be
  T \gg {\Gamma \over g^2} \quad \textrm{(adiabatic)}
\,.
\label{eq:adibound}
\ee
Note that the same quantity appears in the numerator of both expressions;
in both cases, $\Gamma$ accounts for the possibility of transitions to all
possible excited states.

The adiabatic and measurement algorithms have qualitatively similar
behavior: if the gap is exponentially small, neither algorithm is
efficient, whereas if the gap is only polynomially small, both algorithms
are efficient.  However, the measurement algorithm is slightly slower:
whereas adiabatic evolution runs in a time that grows like $1/g^2$, the
measurement algorithm runs in a time that grows like $1/g^3$.  To see that
this comparison is fair, recall that we have defined the momentum in
(\ref{eq:momentum}) so that $||p|| \sim 1$, which gives $||H(s)|| \sim
||H(s) \otimes p||$.  Alternatively, we can compare the number $\eta$ of
few-qubit unitary gates needed to simulate the two algorithms on a
conventional quantum computer.  Using the Lie product formula
\be
  e^{A+B} \simeq (e^{A/m} e^{B/m})^m
\,,
\ee
which is valid provided $m \gg \norm{A}^2+\norm{B}^2$, we find $\eta =
O(1/g^4)$ for adiabatic evolution and $\eta = O(1/g^6)$ for the
measurement algorithm, in agreement with the previous comparison.

The argument we have used to motivate {(\ref{eq:measurebound}) is
explained in greater detail in Appendix~\ref{app:measurement}.  There, we
also consider the number of qubits, $r$, that must be used to represent
the pointer.  We show that if the gap is only polynomially small in $n$,
it is always sufficient to take $r=O(\log n)$.  However, we argue that
generally a single qubit will suffice.

\section{The Grover problem}
\label{sec:grover}

The unstructured search problem considered by Grover is to find a
particular unknown $n$-bit string $w$ (the marked state, or the {\em
winner}) using only queries of the form ``is $z$ the same as $w$?''
\cite{Gro97}.  In other words, one is trying to minimize a function \be
  h_w(z) = \left\{ \matrix{0 \,,\quad z=w \cr 1 \,,\quad z \ne w} \right.
\,.
\ee
Since there are $2^n$ possible values for $w$, the best possible classical
algorithm uses $\Theta(2^n)$ queries.  However, Grover's algorithm
requires only $\Theta(2^{n/2})$ queries, providing a (provably optimal
\cite{BBBV97}) quadratic speedup.  In Grover's algorithm, the winner is
specified by an {\em oracle} $U_w$ with
\be
  U_w |z\> = (-1)^{h_w(z)}|z\>
\,.
\ee
This oracle is treated as a black box that one can use during the
computation.  One call to this black box is considered to be a single
query of the oracle.

In addition to Grover's original algorithm, quadratic speedup can also be
achieved in a time-independent Hamiltonian formulation \cite{FG98a} or by
adiabatic quantum computation \cite{RC01,DMV01}.  In either of these
formulations, the winner is specified by an ``oracle Hamiltonian''
\be
  H_w = 1-|w\>\<w|
\ee
whose ground state is $|w\>$ and that treats all orthogonal states (the
non-winners) equivalently.  One is provided with a black box that
implements $H_w$, where $w$ is unknown, and is asked to find $w$.  Instead
of counting queries, the efficiency of the algorithm is quantified in
terms of the total time for which one applies the oracle Hamiltonian.

Here, we show that if we are given a slightly different black box, we can
achieve quadratic speedup using the measurement algorithm.  We let the
problem Hamiltonian be $H_P=H_w$ and we consider a one-parameter family of
Hamiltonians $H(s)$ given by (\ref{eq:lininterp}) for some $H_B$.  Because
we would like to {\em measure} this Hamiltonian, it is not sufficient to
be given a black box that allows one to evolve the system according to
$H_w$.  Instead, we will use a black box that evolves the system and a
pointer according to $H_w \otimes p$, where $p$ is the momentum of the
pointer.  This oracle is compared to the previous two in
Fig.~\ref{fig:oracle}.  By repeatedly alternating between applying this
black box and evolving according to $H_B \otimes p$, each for small time,
we can produce an overall evolution according to the Hamiltonian $[s
H_B+(1-s)H_P]\otimes p$, and thus measure the operator $H(s)$ for any $s$.

\begin{figure}
\begin{tabular}{l}
\includegraphics[width=125pt]{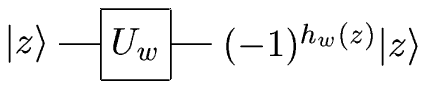} \\ \\
\includegraphics[width=115pt]{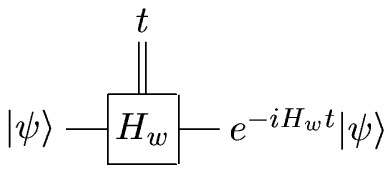} \\ \\
\includegraphics[width=192pt]{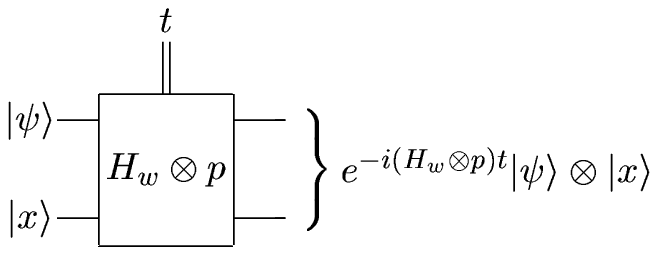}
\end{tabular}
\caption{Oracles for the Grover problem.  (a) Top: Grover's original
oracle.  (b) Center: An oracle that performs evolution according to $H_w$.
The double line indicates a classical control parameter, the time for
which the Hamiltonian is applied.  (c) Bottom: An oracle that allows one
to measure $H_w$.}
\label{fig:oracle}
\end{figure}

Now consider the beginning Hamiltonian
\be
  H_B = \sum_j {1-\sigma_x^{(j)} \over 2}
\,,
\label{eq:HB}
\ee
where $\sigma_x^{(j)}$ is the Pauli $x$ operator acting on the $j$th
qubit.  This beginning Hamiltonian is a sum of local terms, and has the
easy-to-prepare ground state $|E_0(0)\>=2^{-n/2} \sum_z |z\>$, the uniform
superposition of all possible bit strings in the computational basis.  If
we consider the interpolation (\ref{eq:lininterp}), then one can show
\cite{FGGS00} that the minimum gap occurs at
\be
  s^* = 1 - {2 \over n} + O(n^{-2})
\,,
\label{eq:sstar}
\ee
where the gap takes the value
\be
  g(s^*) = 2^{1-n/2}[1+O(n^{-1})]
\,.
\ee
Naively applying (\ref{eq:measurebound}) gives a running time $T =
O(2^{3n/2})$, which is even worse than the classical algorithm.

However, since we know the value of $s^*$ independent of $w$, we can
improve on this approach by making fewer measurements.  We observe that in
the limit of large $n$, the ground state of $H(s)$ is close to the ground
state $|E_0(0)\>$ of $H_B$ for $s \alt s^*$ and is close to the ground
state $|E_0(1)\>=|w\>$ of $H_P$ for $s \agt s^*$, switching rapidly from
one state to the other in the vicinity of $s=s^*$.  In
Appendix~\ref{app:groverground}, we show that up to terms of order $1/n$,
the ground state $|\psi_+\>$ and the first excited state $|\psi_-\>$ of
$H(s^*)$ are the equal superpositions
\be
  |\psi_\pm\> \simeq {1 \over \sqrt 2}(|E_0(0)\> \pm |E_0(1)\>)
\label{eq:groverground}
\ee
of the initial and final ground states (which are nearly orthogonal for
large $n$).  If we prepare the system in the state $|E_0(0)\>$ and make a
perfect measurement of $H(s^*)$ followed by a perfect measurement of
$H(1)$, we find the result $w$ with probability $1 \over 2$.  The same
effect can be achieved with an imperfect measurement, even if the pointer
consists of just a single qubit.  First consider the measurement of
$H(s^*)$ in the state $|E_0(0)\>$.  After the system and pointer have
interacted for a time $t$ according to (\ref{eq:pointerevolve}) with
$r=1$, the reduced density matrix of the system in the
$\{|\psi_+\>,|\psi_-\>\}$ basis is approximately
\be
  {1 \over 2} \left(\matrix{1 & e^{ ig(s^*)t/4} \cos[g(s^*)t/4] \cr
                                e^{-ig(s^*)t/4} \cos[g(s^*)t/4] & 1}\right)
\,.
\ee
If we then measure $H(1)$ (i.e., measure in the computational basis), the
probability of finding $w$ is approximately
\be
  {1 \over 2} \sin^2 [g(s^*)t/4]
\,.
\ee
To get an appreciable probability of finding $w$, we choose
$t=\Theta(2^{n/2})$.

This approach is similar to the way one can achieve quadratic speedup with
the adiabatic algorithm.  Schr\"odinger time evolution governed by
(\ref{eq:lininterp}) does not yield quadratic speedup.  However, because
$s^*$ is independent of $w$, we can change the Hamiltonian quickly when
the gap is big and more slowly when the gap is small.  Since the gap is
only of size $\sim 2^{-n/2}$ for a region of width $\sim 2^{-n/2}$, the
total oracle time with this modified schedule need only be $O(2^{n/2})$.
This has been demonstrated explicitly by solving for the optimal schedule
using a different beginning Hamiltonian $H_B'$ that is not a sum of local
terms \cite{RC01,DMV01}, but it also holds using the beginning Hamiltonian
(\ref{eq:HB}).

Note that measuring $H(s^*)$ is not the only way to solve the Grover
problem by measurement.  More generally, we can start in some
$w$-independent state, measure the operator
\be
  \tilde H = H_w + K
\label{eq:generalgrover}
\ee
where $K$ is also independent of $w$, and then measure in the
computational basis.  For example, suppose we choose 
\be
  K = 1 - |\psi\>\<\psi|
\,,
\label{eq:groverk}
\ee
where $|\psi\>$ is a $w$-independent state with the property $|\<w|\psi\>|
\sim 2^{-n/2}$ for all $w$.  (If we are only interested in the time for
which we use the black box shown in Fig.~\ref{fig:oracle}(c), i.e., if we
are only interested in the oracle query complexity, then we need not
restrict $K$ to be a sum of local terms.)   In (\ref{eq:groverk}), the
coefficient of $-1$ in front of $|\psi\>\<\psi|$ has been fine-tuned so
that $|\psi\>+|w\>$ is the ground state of $\tilde H$ (choosing the phase
of $|w\>$ so that $\<w|\psi\>$ is real and positive).  If the initial
state has a large overlap with $|\psi\>$, then the measurement procedure
solves the Grover problem.  However, the excited state $|\psi\>-|w\>$ is
also an eigenstate of $\tilde H$, with an energy higher by of order
$2^{-n/2}$.  Thus the time to perform the measurement must be
$\Omega(2^{n/2})$.

The measurement procedures described above saturate the well-known lower
bound on the time required to solve the Grover problem.  Using an oracle
like the one shown in Fig.~\ref{fig:oracle}(a), Bennett {\em et al.}\ 
showed that the Grover problem cannot be solved on a quantum computer
using fewer than of order $2^{n/2}$ oracle queries \cite{BBBV97}.  By a
straightforward modification of their argument, an equivalent result
applies using the oracle shown in Fig.~\ref{fig:oracle}(c).  Thus every
possible $\tilde H$ as in (\ref{eq:generalgrover}) that can be measured to
find $w$ must have a gap between the energies of the relevant eigenstates
of order $2^{-n/2}$ or smaller.

\section{Discussion}
\label{sec:discussion}

We have described a way to solve combinatorial search problems on a
quantum computer using only a sequence of measurements to keep the
computer near the ground state of a smoothly varying Hamiltonian.  The
basic principle of this algorithm is similar to quantum computation by
adiabatic evolution, and the running times of the two methods are closely
related.  Because of this close connection, many results on adiabatic
quantum computation can be directly imported to the measurement algorithm
--- for example, its similarities and differences with classical simulated
annealing \cite{FGG02}.  We have also shown that the measurement algorithm
can achieve quadratic speedup for the Grover problem using knowledge of
the place where the gap is smallest, as in adiabatic quantum computation.  

One of the advantages of adiabatic quantum computation is its inherent
robustness against error \cite{CFP02}.  In adiabatic computation, the
particular path from $H_B$ to $H_P$ is unimportant as long as the initial
and final Hamiltonians are correct, the path is smoothly varying, and the
minimum gap along the path is not too small.  Exactly the same
considerations apply to the measurement algorithm.  However, the adiabatic
algorithm also enjoys robustness against thermal transitions out of the
ground state: if the temperature of the environment is much smaller than
the gap, then such transitions are suppressed.  The measurement algorithm
might not possess this kind of robustness, since the Hamiltonian of the
quantum computer during the measurement procedure is not simply $H(s)$.

Although it does not provide a computational advantage over quantum
computation by adiabatic evolution, the measurement algorithm is an
alternative way to solve general combinatorial search problems on a
quantum computer.  The algorithm can be simply understood in terms of
measurements of a set of operators, without reference to unitary time
evolution.  Nevertheless, we have seen that to understand the running time
of the algorithm, it is important to understand the dynamical process by
which these measurements are realized.

\acknowledgments

This work was supported in part by the Department of Energy under
cooperative research agreement DE-FC02-94ER40818, by the National Science
Foundation under grant EIA-0086038, and by the National Security Agency
and Advanced Research and Development Activity under Army Research Office
contract DAAD19-01-1-0656.  AMC received support from the Fannie and John
Hertz Foundation, and ED received support from the Istituto Nazionale di
Fisica Nucleare (Italy).

\appendix
\section{The measurement process}
\label{app:measurement}

In Section \ref{sec:runtime}, we discussed the running time of the
measurement algorithm by examining the measurement process.  In this
Appendix, we present the analysis in greater detail.  First, we derive the
bound on the running time by demonstrating (\ref{eq:numbermeas}) and
(\ref{eq:meastime}).  We show rigorously that these bounds are sufficient
as long as the gap is only polynomially small and the number of qubits
used to represent the pointer is $r=O(\log n)$.  Finally, we argue that
$r=1$ qubit should be sufficient in general.

Our goal is to find a bound on the final success probability of the
measurement algorithm.  We consider the effect of the measurements on the
reduced density matrix of the system, which can be written as the block
matrix
\be
  \rho = \left( \matrix{ \mu & \nu^\dag \cr
                         \nu & \chi } \right)
\ee
where $\mu = \rho_{00}$, $\nu_a = \rho_{a0}$ for $a \ne 0$, and
$\chi_{ab}=\rho_{ab}$ for $a,b \ne 0$.  Since $\tr\rho=1$, $\mu = 1 -
\tr\chi$.  For ease of notation, we suppress $j$, the index of the
iteration, except where necessary.  The unitary transformation
(\ref{eq:unitary}) may also be written as a block matrix.  Define
$\epsilon = \Gamma \delta / g$.  Using perturbation theory and the
unitarity constraint, we can write
\be
  U = \left( \matrix{ u & -w^\dag V + O(\epsilon^3) \cr
                      w & V + O(\epsilon^2) } \right)
\,,
\ee
where $|u|^2 \ge 1 - \epsilon^2 + O(\epsilon^3)$, $\norm{w}^2 \le
\epsilon^2 + O(\epsilon^3)$, and $V$ is a unitary matrix.  We let
$\norm\cdot$ denote the $l_2$ vector or matrix norm as appropriate.
Furthermore, let
\be
  \kappa = \left( \matrix{ 1 & k^\dag \cr k & J} \right)
\,.
\ee
From (\ref{eq:systempointer}), the effect of a single measurement may be
written
\be
  \rho' = (U \rho U^\dag) \circ \kappa
\,,
\ee
where $\circ$ denotes the element-wise (Hadamard) product.  If we assume
$\norm\nu = O(\epsilon)$, we find
\bea
  \label{eq:mu}
  \mu'  &=& |u|^2 \mu - w^\dag V \nu - \nu^\dag V^\dag w + O(\epsilon^3) \\
  \nu'  &=& [V \nu + \mu w - V \chi V^\dag w + O(\epsilon^2)] \circ k
  \label{eq:nu}
\,.
\eea

Now we use induction to show that our assumption always remains valid.
Initially, $\nu^{(0)}=0$.  Using the triangle inequality in
(\ref{eq:nu}), we find
\be
  \norm{\nu'} \le [\norm\nu + \epsilon + O(\epsilon^2)] \tilde k
\,,
\ee
where
\be
  \tilde k = \max_{j,a} \left|k^{(j)}_a\right|
\,.
\ee
So long as $\tilde k < 1$, we can sum a geometric series, extending the
limits to go from $0$ to $\infty$, to find
\be
  \norm{\nu^{(j)}} \le {\epsilon \over 1-\tilde k} + O(\epsilon^2)
\ee
for all $j$.  In other words, $\norm\nu = O(\epsilon)$ so long as $\tilde
k$ is bounded below 1 by a constant.

Finally, we put a bound on the final success probability $\mu^{(M)}$.
Using the Cauchy-Schwartz inequality in (\ref{eq:mu}) gives
\be
  \mu' \ge (1-\epsilon^2)\mu - {2\epsilon^2 \over 1 - \tilde k} 
           + O(\epsilon^3)
\,.
\ee
Iterating this bound $M$ times with the initial condition $\mu^{(0)}=1$,
we find
\be
  \mu^{(M)} \ge 1 - {\Gamma^2 \over M g^2} 
                \left(1 + {2 \over 1 - \tilde k}\right)
                + O(M \epsilon^3)
\,.
\ee
If $\tilde k$ is bounded below 1 by a constant (independent of $n$), we
find the condition (\ref{eq:numbermeas}) as claimed in Section
\ref{sec:runtime}.

The requirement on $\tilde k$ gives the bound (\ref{eq:meastime}) on the
measurement time $\tau$, and also gives a condition on the number of
pointer qubits $r$.  To see this, we must investigate properties of the
function $|\kappa(x)|^2$ defined in (\ref{eq:kappa}) and shown in
Fig.~\ref{fig:kappa}.  It is straightforward to show that $|\kappa(x)|^2
\le 1/2$ for $\pi/2 \le x \le \pi(2^r-1/2)$.  Thus, if we want $\tilde k$
to be bounded below 1 by a constant, we require
\be
  \pi/2 \le [E_a(s) - E_0(s)] t/2 \le \pi(2^r-1/2)
\label{eq:window}
\ee
for all $s$ and for all $a \ne 0$.  The left hand bound with $a=1$ gives
$t \ge \pi/g$, which is (\ref{eq:meastime}).  Requiring the right hand
bound to hold for the largest energy difference gives the additional
condition $2^r \agt (E_{2^n-1}-E_0)/g$.  Since we only consider
Hamiltonians $H(s)$ that are sums of $\poly(n)$ terms of bounded size, the
largest possible energy difference must be bounded by a polynomial in $n$.
If we further suppose that $g$ is only polynomially small, this condition
is satisfied by taking
\be
  r = O(\log n)
\,,
\label{eq:pointersize}
\ee
as claimed at the end of Section \ref{sec:runtime}.  Thus we see that the
storage requirements for the pointer are rather modest.

However, the pointer need not comprise even this many qubits.  Since the
goal of the measurement algorithm is to keep the system close to its
ground state, it would be surprising if the energies of highly excited
states were relevant.  Suppose we take $r=1$; then $|\kappa(x)|^2 =
\cos^2(x/2)$.  As before, (\ref{eq:meastime}) suffices to make
$|\kappa_{01}|^2$ sufficiently small.  However, we must also consider
terms involving $|\kappa_{0a}|^2$ for $a>1$.  The algorithm will fail if
the term $\mu w \circ k$ in (\ref{eq:nu}) accumulates to be $O(1)$ over
$M$ iterations.  This will only happen if, for $O(M)$ iterations, most of
$\norm w$ comes from components $w_a$ with $(E_a-E_0)t$ close to an
integer multiple of $2\pi$.  In such a special case, changing $t$ will
avoid the problem.  An alternative strategy would be to choose $t$ from a
random distribution independently at each iteration.

\section{Eigenstates in the Grover problem}
\label{app:groverground}

Here, we show that the ground state of $H(s^*)$ for the Grover problem is
close to (\ref{eq:groverground}).  Our analysis follows Section 4.2 of
\cite{FGGS00}.

Since the Grover problem is invariant under the choice of $w$, we consider
the case $w=0$ without loss of generality.  In this case, the problem can
be analyzed in terms of the total spin operators
\be
  S_a = {1 \over 2} \sum_{j=1}^n \sigma_a^{(j)}
\,,
\ee
where $a=x,y,z$ and $\sigma_a^{(j)}$ is the Pauli $a$ operator acting on
the $j$th qubit.  The Hamiltonian commutes with $\vec S^2 =
S_x^2+S_y^2+S_z^2$, and the initial state has $\vec S^2 = {n \over 2}({n
\over 2}+1)$, so we can restrict our attention to the $(n+1)$-dimensional
subspace of states with this value of $\vec S^2$.  In this subspace, the
eigenstates of the total  spin operators satisfy
\be
  S_a|m_a=m\> = m|m_a=m\>
\ee
for $m=-{n \over 2},-{n \over 2}+1,\ldots,{n \over 2}$.
Written in terms of the total spin operators and eigenstates, the
Hamiltonian is
\bea
  H(s) &=& (1-s) \left({n \over 2}-S_x\right)  \nonumber\\
        &&   + s \left(1 - \left|m_z = {n \over 2}\right\>
	         \left\<m_z = {n \over 2}\right|\right)
\,.
\eea
The initial and final ground states are given by $|E_0(0)\> = |m_x = {n
\over 2}\>$ and $|E_0(1)\> = |m_z = {n \over 2}\>$, respectively.

Projecting the equation $H(s)|\psi\>=E|\psi\>$ onto the eigenbasis of
$S_x$, we find
\be
  \left\<m_x = {n \over 2}-r \Big| \psi\right\> 
  = {s \over 1-s}{\sqrt{P_r} \over r-\lambda}
    \left\<m_z = {n \over 2} \Big| \psi \right\>
\,,
\label{eq:eigen}
\ee
where we have defined $\lambda = (E-s)/(1-s)$ and $P_r = 2^{-n} {n \choose
r}$.  Now focus on the ground state $|\psi_+\>$ and the first excited
state $|\psi_-\>$ of $H(s^*)$.  By equation (4.39) of \cite{FGGS00}, these
states have $\lambda_\pm = \mp {n \over 2}2^{-n/2}(1+O(1/n))$.  Putting
$r=0$ in (\ref{eq:eigen}) and taking $s=s^*$ from (\ref{eq:sstar}), we
find
\be
  \left\<m_x = {n \over 2} \Big| \psi_\pm \right\> 
  = \pm \left\<m_z = {n \over 2} \Big| \psi_\pm \right\> (1+O(1/n))
\,.
\label{eq:xzsame}
\ee
For $r \ne 0$, we have
\bea
  \left\<m_x = {n \over 2}-r \Big| \psi_\pm \right\> 
  &=& {n \over 2} {\sqrt{P_r} \over r} 
    \left\<m_z = {n \over 2} \Big| \psi_\pm \right\> \nonumber\\
  && \times (1+O(1/n))
\,.
\eea
Requiring that $|\psi_\pm\>$ be normalized, we find
\bea
  1 &=& \sum_{r=0}^n \left|\left\<m_x = {n \over 2}-r 
        \Big| \psi_\pm\right\>\right|^2  \\
    &=& \left|\left\<m_z = {n \over 2} \Big| \psi_\pm \right\>\right|^2 
        \left(1 + {n^2 \over 4} \sum_{r=1}^n {P_r \over r^2}\right) \nonumber\\
     && \times (1+O(1/n)) \\
    &=& \left|\left\<m_z = {n \over 2} \Big| \psi_\pm \right\>\right|^2
        (2+O(1/n))
\,,
\eea
which implies $|\<m_z = {n \over 2}|\psi_\pm\>|^2 = {1 \over 2}+O(1/n)$.
From (\ref{eq:xzsame}), we also have $|\<m_x = {n \over 2}|\psi_\pm\>|^2 =
{1 \over 2}+O(1/n)$.  Thus we find
\be
  |\psi_\pm\> \simeq {1 \over \sqrt2} \left( \left|m_x={n \over 2}\right\> 
                                  \pm \left|m_z={n \over 2}\right\> \right)
\ee
up to terms of order $1/n$, which is (\ref{eq:groverground}).



\begin{thebibliography}{}
\bibitem{Sho94}
  P. W. Shor, {\em Algorithms for quantum computation: discrete logarithms
  and factoring}, in Proc. 35th Annual Symposium on Foundations of
  Computer Science, ed. S. Goldwasser, 124 (IEEE Press, Los Alamitos, CA,
  1994).
\bibitem{Gro97}
  L. K. Grover, {\em Quantum mechanics helps in searching for a needle in
  a haystack}, Phys. Rev. Lett. {\bf 79}, 325 (1997).
\bibitem{FGGS00}
  E. Farhi, J. Goldstone, S. Gutmann, and M. Sipser, {\em Quantum
  computation by adiabatic evolution}, quant-ph/0001106.
\bibitem{Nie01}
  M. A. Nielsen, {\em Universal quantum computation using only projective
  measurement, quantum memory, and preparation of the 0 state},
  quant-ph/0108020.
\bibitem{FZ01}
  S. A. Fenner and Y. Zhang, {\em Universal quantum computation with
  two- and three-qubit projective measurements}, quant-ph/0111077.
\bibitem{Leu01}
  D. W. Leung, {\em Two-qubit projective measurements are universal for
  quantum computation}, quant-ph/0111122.
\bibitem{RB00}
  R. Raussendorf and H. J. Briegel, {\em Quantum computing via
  measurements only}, quant-ph/0010033.
\bibitem{Neu32}
  J. von Neumann, {\em Mathematical Foundations of Quantum Mechanics}
  (Princeton University Press, Princeton, NJ, 1955).
\bibitem{AV80}
  Y. Aharonov and M. Vardi, {\em Meaning of an individual ``Feynman
  path,''} Phys. Rev. D {\bf 21}, 2235 (1980).
\bibitem{SRM93}
  L. S. Schulman, A. Ranfagni, and D. Mugnai, {\em Characteristic scales
  for dominated time evolution}, Physica Scripta {\bf 49}, 536 (1993).
\bibitem{CFGG01}
  A. M. Childs, E. Farhi, J. Goldstone, and S. Gutmann, {\em Finding
  cliques by quantum adiabatic evolution}, quant-ph/0012104.
\bibitem{Wie96}
  S. Wiesner, {\em Simulations of many-body quantum systems by a quantum
  computer}, quant-ph/9603028.
\bibitem{Zal98}
  C. Zalka, {\em Simulating quantum systems on a quantum computer}, Proc.
  R. Soc. London A {\bf 454}, 313 (1998).
\bibitem{Llo96}
  S. Lloyd, {\em Universal quantum simulators}, Science {\bf 273}, 1073
  (1996).
\bibitem{Kit95}
  A. Yu. Kitaev, {\em Quantum measurements and the abelian stabilizer
  problem}, quant-ph/9511026.
\bibitem{CEMM98}
  R. Cleve, A. Ekert, C. Macchiavello, and M. Mosca, {\em Quantum
  algorithms revisited}, Proc. R. Soc. London A {\bf 454}, 339 (1998).
\bibitem{AL99}
  D. S. Abrams and S. Lloyd, {\em Quantum algorithm providing exponential
  speed increase for finding eigenvalues and eigenvectors}, Phys. Rev.
  Lett. {\bf 83}, 5162 (1999).
\bibitem{TMR02}
  B. C. Travaglione, G. J. Milburn, and T. C. Ralph, {\em Phase estimation
  as a quantum nondemolition measurement}, quant-ph/0203130.
\bibitem{BBBV97}
  C. H. Bennett, E. Bernstein, G. Brassard, and U. Vazirani, {\em
  Strengths and weaknesses of quantum computing}, SIAM J. Comput. {\bf
  26}, 1510 (1997).
\bibitem{FG98a}
  E. Farhi and S. Gutmann, {\em Analog analogue of a digital quantum
  computation}, Phys. Rev. A {\bf 57}, 2403 (1998).
\bibitem{RC01}
  J. Roland and N. Cerf, {\em Quantum search by local adiabatic
  evolution}, quant-ph/0107015.
\bibitem{DMV01}
  W. van Dam, M. Mosca, and U. Vazirani, {\em How powerful is adiabatic
  quantum computation?}, to appear in FOCS 2001.
\bibitem{FGG02}
  E. Farhi, J. Goldstone, and S. Gutmann, {\em Quantum adiabatic evolution
  algorithms versus simulated annealing}, quant-ph/0201031.
\bibitem{CFP02}
  A. M. Childs, E. Farhi, and J. Preskill, {\em Robustness of adiabatic
  quantum computation}, Phys. Rev. A {\bf 65}, 012322 (2002).
\end{thebibliography}
\end{document}